# Delayed formation of coherent LO phonon-plasmon coupled modes in n-type and p-type GaAs measured using a femtosecond coherent control technique

Jianbo Hu,[1,2,4] Oleg V. Misochko,[3] Arihiro Goto,[1,2] Kazutaka G. Nakamura[1,2]

[1]*Materials and Structures Laboratory, Tokyo Institute of Technology, R3-10, 4259 Nagatsuta, Yokohama 226-8503, Japan*

[2]*CREST, Japan Science and Technology Agency, Kawaguchi, Saitama 332-0012, Japan*

[3]*Institute of Solid State Physics, Russian Academy of Sciences, 142432 Chernogolovka, Moscow region, Russia*

[4]*Institute of Fluid Physics, China Academy of Engineering Physics, Mianyang 621900, Sichuan, China*

Coherent control experiments using a pair of collinear femtosecond laser pulses have been carried out to manipulate longitudinal optical (LO) phonon-plasmon coupled (LOPC) modes in both p- and n-type GaAs. By tuning the interpulse separation, remarkably distinct responses have been observed in the two samples. To understand the results obtained a phenomenological model taking the delayed formation of coherent LOPC modes into account is proposed. The model suggests that the lifetime of coherent LOPC modes plays a key role and the interference of the coherent LO phonons excited successively by two pump pulses strongly affects the manipulation of coherent LOPC modes.

# I. INTRODUCTION

In polar semiconductors when the plasmon and the longitudinal optical (LO) phonons have comparable frequencies the two longitudinal modes become mixed to generate the LO phonon-plasmon coupled (LOPC) mode [1,2]. The dynamics of the LOPC mode are well described by a pair of coupled equations for the electronic polarization and the normalized lattice displacement [3,4]. For GaAs, one of the most studied polar semiconductors, both the frequency- [3, 5-7] and time-domain [8-12] experiments demonstrated that, depending on the carrier type, the LOPC mode could exhibit considerably varying properties. The LO phonon-hole plasmon coupled mode has only single branch with the frequency between the LO and transverse-optical (TO) phonons [6, 7, 11, 12]. This peculiar behavior is due to overdamped character of hole plasmon [6, 7]. The LO phonon-electron plasmon coupled mode, nevertheless, splits into two anti-crossing branches: the upper branch mode $L_+$ and the lower branch mode $L_-$ [3].

Although numerous experiments verified the validity of the coupled equations, one may notice that the equations simply assumed an instantaneous coupling between two longitudinal modes. However, recent laser pump-THz probe experiments clearly indicate that the formation of the plasmon in GaAs is actually time-consuming [13]. The transition from the photoexcited electron-hole plasma to collective response of the many-body system takes around one period of the plasmon (less than 100 fs). Moreover, both theories and experiments of quantum kinetics have demonstrated that the LOPC modes are not established instantaneously and the coupling of two longitudinal modes is a kinetic process controlled by time and carrier density [14-16]. It is not surprising, however, that the delayed formation of coupled modes has no considerable influence on the observed dynamics because most of the frequency-domain measurements are incapable to resolve such short time scale. The time-domain measurements with the single pump-pulse are able to resolve such kinetics but usually ignore it, because such ultrafast dynamics only affects the initial phase of coherently excited modes [17].

On the other hand, coherent control experiments in p-GaAs [18] showed an interesting feature. Namely, the amplitude modulation of the $L_-$ mode as a function of the interpulse separation was synchronized to that of coherent LO phonons. This result is in a sharp contrast to the case of n-GaAs in which the amplitude of the $L_-$ mode was modulated independently of the phonon mode [19]. To explain the peculiarity of coherent control Ref. [18] suggested a possible influence of the delayed formation on coherent control. It is also worth noting that a theoretical work tried to describe coherent control with two pump pulses [20]. Based on the coupled equations Bukunov *et al.* assumed the same procedure for the second pump excitation as for the first one, and succeeded in simulating coherent control experiments in n-GaAs. But the results of coherent control in p-GaAs cannot be well reproduced by this model.

In our paper we demonstrate the manipulation of coherent LO phonons and LOPC modes in both p- and n-type GaAs using a pair of temporally spaced femtosecond laser pulses. A simple phenomenological model is proposed to interpret different results of coherent control in the two samples.

## II. EXPERIMENTAL

The samples used were a Zn-doped p-GaAs(100) with the carrier density of $N=5\times10^{18}$ cm$^{-3}$ and a Si-doped n-GaAs(100) with the carrier density of $N=5\times10^{17}$ cm$^{-3}$. Both crystals obtained from MTI Corporation were grown by the liquid encapsulated Czochralski method. To study the ultrafast lattice and carrier dynamics the femtosecond reflection-type pump-probe technique with electro-optic detection was employed [8]. Ultrashort laser pulses with the duration of 45 fs and the photon energy of 1.55 eV ($\lambda=800$ nm) acted as a pump and a probe, the polarizations of which were set orthogonal to each other to eliminate the coherent artifact. The pulses were delivered by a Kerr-lens mode-locked Ti: sapphire oscillator at the repetition rate of around 80 MHz. Excitation was achieved slightly above the direct gap (~1.44 eV) at Γ-point. The delay line was modulated by a fast scan delay at the frequency of 20 Hz to obtain the transient reflection change $\Delta R_{eo}(t)/R_0$, where $\Delta R_{eo}(t)$ is the difference between two orthogonal components of the reflected probe with pump, and $R_0$ the reflectivity without pump. To perform coherent control experiments, the pump pulse was fed into a modified Mach-Zehnder interferometer to be divided into two collinear beams with a variable interpulse separation, Δt, controlled by a translation stage. All the measurements were carried at room temperature with the probe fluence of ~3 μJ/cm².

## III. RESULTS AND DISCUSSION

### A. Single-pump excitation

In n-GaAs, the time-domain signal shown in Fig. 1(a) demonstrates a clear mode beating. In p-GaAs, although the beating feature is not that obvious as in n-GaAs, one can easily see that the ultrafast oscillations shown in Fig. 1(b) undergo successively fast and slow decay, in which the fast decay lasts only a few hundreds of femtoseconds. The Fourier transformed (FT) spectra, shown in the insets of Fig. 1, confirm that in both samples two modes are coherently excited, one of which has the frequency $\omega_L/2\pi = 8.7$ THz corresponding to the LO phonon. The other mode in both samples has a frequency smaller than that of the TO phonon (8.1 THz). We assign this mode to the lower branch, L₋, of the LO phonon-electron plasmon coupled modes. Such two-mode behavior can originate from the spatial (either lateral or depth) inhomogeneity of the photoexcited carrier density induced by ultrashort laser pulses, whereas the absence of upper branch $L_+$ mode might be due to overdamping. To obtain the mode parameters for the oscillations shown in Fig. 1, we modeled the transient reflectivity by two damped harmonic oscillators. Note that the FT spectra for fitted and experimental signals are just the same. Interestingly, the fitted lifetime of the $L_-$ mode in n-GaAs is about 5 times larger than that in p-GaAs, which can be due to ultrafast carrier scattering in the latter sample [21]. The two similarly excited modes with significantly different lifetimes provide us a good opportunity to make a comparison study about the role of lifetime in coherent control.

The frequency of the $L_-$ mode depends on the total carrier density including background and photo-doped carriers [9]. Therefore, we further examined the mode assignment through measuring its dependence on the photoexcited carrier density. To this end, we varied the pump fluence from 19 µJ/cm$^2$ to 122 µJ/cm$^2$, corresponding to the photoexcited carrier density from $6.48 \times 10^{17}$ cm$^{-3}$ to $4.15 \times 10^{18}$ cm$^{-3}$, to detect the transient signal at different excitation strengths. The Fourier transformed spectra and the fitted parameters of the time-domain signals are plotted as a function of the pump fluence in Fig. 2 and Fig. 3, respectively. In both samples we observed that the frequency of the $L_-$ mode substantially depends on the pump fluence. With the increasing pump fluence (or photoexcited carrier density) the frequency of the $L_-$ mode tends to that of TO phonon. Therefore, the $L_-$ mode at high excitation strength is phononlike. Its lifetime is almost independent of the pump fluence. In contrast, the lifetime of LO phonon monotonically decreases with the increasing pump fluence, whereas its frequency remains essentially unchanged. We attribute the decreasing lifetime to the increasing phonon-phonon and phonon-carrier interactions. The amplitudes of two modes in the two samples, in the experimentally available range of the pump fluence, exhibit an opposite trend on the pump fluence. In p-GaAs, the saturation of LO phonons comes from the complete screening of the built-in field, whereas the almost linear rise of the L_ mode suggests that its increase is due to the plasmon contribution. In n-GaAs, the amplitude of LO phonons linearly increases with increasing pump fluence, thus suggesting an incomplete screening, whereas in p-GaAs the amplitude tends to saturate. The varying response of LO phonons in the two samples in the same range of pump fluence can be due to different built-in fields, which is higher in n-GaAs than in p-GaAs [8]. The sublinear rise of the L_ mode, the same as observed by Ishioka *et al*. [11], indicates its direct relation to the photoexcited carrier density. In addition, we noticed a phase difference of ~$\pi/2$ between two coherent modes, shown in Fig. 4, which may be indicative of kinetic process responsible for the generation of LOPC modes [17].

**B. Double-pump excitation**

As mentioned above, while the dynamics of coupled modes in the frequency-domain, as well as in the time-domain with a single-pump excitation, are relatively well understood [3, 5-12], the prospect of manipulating their excitation with two ultrashort laser pulses has not yet been fully addressed. To achieve coherent control, we used two collinear pump pulses to excite and control the ultrafast dynamics of coupled modes. In n-GaAs, the photoexicted carrier densities induced by every pump pulse were $1.94 \times 10^{18}$ cm$^{-3}$ and $1.43 \times 10^{18}$ cm$^{-3}$. In p-GaAs, the carrier densities equaled to $3.77 \times 10^{18}$ cm$^{-3}$ and $3.56 \times 10^{18}$ cm$^{-3}$.

We varied the interpusle separation from 315 fs to 545 fs. In both samples the delay between two pump pulses affects the time-domain signal to a significant degree. To obtain the parameters of the resulting ultrafast response we set the second pump excitation as the zero time and fitted the time-domain signal. The fitted mode parameters as a function of the interpulse separation are presented in Fig. 5. There are two interesting features in the results obtained. First, in n-GaAs, the amplitudes of the LO phonons and L_ mode are harmonically modulated by the second pump pulse. By fitting the data by an exponentially damped sine function the modulation periods of two modes are determined to be 133 fs and 114 fs, respectively, corresponding to $L_-$ mode and LO phonon frequencies. This result is consistent with Mizoguchi *et al.*'s observation [19]. In p-GaAs, however, the amplitude of the $L_-$ mode is modulated in phase with that of the LO phonons at the period of LO phonons, although these two modes actually have different frequencies. This observation similar to Ref. [18] in term of modulation period, however, is in striking contrast to Ref. [18] in term of modulation phase. Namely, in Ref. [18] out-of-phase modulation of the two modes was observed, whereas in our case the modulation is in-phase. The disparity might be due to different experimental techniques used to study the ultrafast dynamics. In Ref. [18] a second-harmonic generation pump-probe technique was used, in which either excitation or probe was mainly limited by the depletion layer. Such a limitation may result in somehow different coupling dynamics [11].

The second interesting feature is that the lifetime of the $L_-$ mode in both samples is also modulated by the second pump pulse. Especially in p-GaAs the change in lifetime is quite large exceeding 50 %. These results are not consistent with the pump fluence measurements, in which the lifetime of the $L_-$ mode is not sensitive to the carrier density. On the other hand, one can also expect that tuning the interpulse separation to such short time interval the total carrier density excited by two pump pulses would not be changed; therefore, the modulation of the lifetime of the $L_-$ mode as a function of the interpulse separation is unexpected.

## C. Phenomenological model

It is well known that the generation of coherent LO phonons in GaAs is based on the ultrafast screening of the build-in field in the depletion layer, which takes more than 10 fs [8]. Similar excitation mechanism can be also applicable to the collective carrier excitation (plasmon) [22, 23], but it may take longer time than that for coherent LO phonons. As observed by Huber *et al.* using THz technique, the formation of plasmon takes around one intrinsic period determined by the carrier density [13]. According to our excitation conditions, the estimated formation time of the plasmon is expected to be longer than 50 fs. Note that the coherent excitation of plasmons in our case is somehow different from that in Ref. [13]. In Ref. [13] the coherent excitation is achieved via the (quantum kinetic) buildup of the microscopic Coulomb screening, while in this case that is mainly realized via the (semiclassical) macroscopic screening of the depletion fields associated with the sample surface. Clearly, this is connected to transport of photogenerated charge carriers and will have a finite internal dynamics as well. Taking into account the possible delay of the coupling between the coherent LO phonon and the plasmon after their formations, we expect the formation of the LOPC mode can take even longer time. Therefore, the possible steps in temporal order after the first-pump excitation are as follows: coherent LO phonons are firstly launched, then the plasmon is formed, and finally the two longitudinal modes become coupled (given by Eqs. (1) and (2) in the following paragraph). For the delayed second pump excitation, in principle, a similar kinetic process should occur. Due to the delayed formation of coupled mode, however, it is possible that the coherent LO phonons excited by the second pump pulse interfere with those excited by the first one, before coupling with plasmon (given by Eqs. (3) and (4)). After this interference the coherent LO phonons are coupled to the plasmon to form the LOPC mode, which further interferes with the coupled mode excited by the first pump pulse (given by Eqs. (5) and (6)). The temporal order of the events, schematically shown in Fig. 6, helps in understanding of coherent control experiments in GaAs. Note that our treatment is essentially different from the idea adopted by Bakunov *et al.* [20], in which the authors assumed the same process for each pump pulse to address coherent control.

Based on the idea of the delayed formation, we suggest a phenomenological model to interpret coherent control experiments. From the previous analysis, we know that the dynamics of two coherent modes can be well described by the damped harmonic functions. Therefore, immediately after the first pump excitation, the dynamics of two coherent modes are given by

$$A_{LOPC}^{1}(t) = \alpha \eta A_1 \exp(-(t-\delta)/\tau_{LOPC}) \cos(\omega_{LOPC}(t-\delta)) \tag{1}$$

$$A_{LO}^{1}(t) = A_1(1-\alpha) \exp(-t/\tau_{LO}) \cos(\omega_{LO} t) \tag{2}$$

After the second pump excitation at t=Δt, the dynamics are described by

$$A_{LO}^{2'}(t) = A_2 \exp(-(t-\Delta t)/\tau_{LO}) \cos(\omega_{LO}(t-\Delta t)) \tag{3}$$

$$A_{LO}^{2''}(\delta + \Delta t) = A_{LO}^{1}(\delta + \Delta t) + A_{LO}^{2'}(\delta) \tag{4}$$

$$A_{LOPC}^{2}(\delta + \Delta t) = \alpha \eta A_{LO}^{2''}(\delta + \Delta t) + A_{LOPC}^{1}(\delta + \Delta t) \tag{5}$$

$$A_{LO}^{2}(\delta + \Delta t) = A_{LO}^{2''}(\delta + \Delta t)(1-\alpha) \tag{6}$$

where $A$ is the oscillatory amplitude, $\delta$ the delay time to form the LOPC mode, $\tau$ the lifetime, $\alpha$ the portion of coherent LO phonons coupled to the plasmon, $\eta$ the amplitude ratio of two modes. In this model, all parameters except $\delta$ and $\alpha$ can be directly obtained from the pump dependence. By adjusting the values of $\delta$ and $\alpha$, the main features of the amplitude modulation of the L_ mode and LO phonons in both samples can be well reproduced by this model, as shown in Fig. 5. From this simulation it is easy to see that the lifetime of the L_ mode plays a key role to understand different results in the two samples. Indeed, in p-GaAs the L_ mode has a much shorter lifetime. Therefore, when the second pump pulse arrives at crystal the L_ mode excited by the first pump pulse has already significantly decayed. Thus the L_ mode excited by the second pump pulse has nearly no interference with the first one, and the dynamics are determined mainly by the interference of the coherent LO phonons excited by two pump pulses. In contrast, in n-GaAs the L_ modes excited by each pump pulse experience strong interference due to a longer lifetime. As a result, in this case we can observe the independent modulation of the L_ mode. The formation time $\delta$ obtained from the fit is 110 fs for n-GaAs and 120 fs for p-GaAs, which is close to the period of LO phonon. It is worth mentioning that in our model only coherent LO phonons are coupled to the plasmon, and the calculated results are good enough to explain the experiments. Thermal LO phonons in GaAs do not play a considerable role in controlling the coherent L_ mode. The observed lifetime modulation by the second pump pulse is always anti-correlated with the amplitude modulation in both samples. In Ref. [24], Hase *et al*. ascribed the lifetime increase of coupled modes to the decreased Frohlich-type phonon-carrier interaction. Such explanation does not look suitable for our observation. Actually, the lifetime modulation could be well understood according to the theory and experiments of Vallee *et al*. [25]. For a phononlike coupled mode, the plasmon character strongly affects its lifetime. In our case, it is possible that the lifetime decrease is due to the increase of plasmon contribution when the amplitude of the L_ mode increases, and vice versa.

## IV. CONCLUSIONS

In conclusion, a phenomenological model has been proposed to interpret the results of coherent control experiments in both n- and p- GaAs. Based on the comparison between the experiments and suggested model, the coupling dynamics of the plasmon and coherent LO phonons has been understood, in which the formation of the L_ mode takes around one period of LO phonons, and the lifetime of the L_ mode plays an important role in explaining the substantially different behaviors observed in the two samples.

## ACKNOWLEDGMENTS

We are grateful to Prof. Dr. Alfred Leitenstorfer for helpful and critical discussions .

**Figure captions**

**Fig. 1**. (color online) The oscillatory part of transient reflectivity for the single pump excitation measurement in n-GaAs (a) and p-GaAs (b). Here we have removed the incoherent signal related to the fast generation and slow recombination of photo-injected carriers. The FT spectra of the time-domain signals are shown in the insets. The fitting results (red thin lines) are also presented to compare with the experimental results (green thick lines).

**Fig. 2**. (color online) Fourier transformed spectrum as a function of pump fluence in n-GaAs (a) and p-GaAs (b). The fitting results (blue solid lines) are presented to compare with the experimental results (green dashed lines).

**Fig. 3**. (color online) Amplitude and lifetime of LO (a) and $L_-$ (b) modes as a function of pump fluence in n- and p-GaAs. The diamonds and circles indicate data in n-GaAs and p-GaAs, respectively. The dashed lines are just a guide to the eye.

**Fig. 4**. (Color online) Initial phase of coherent oscillations of LO phonons and $L_-$ modes in n- (a) and p-GaAs (b). The open and filled circles indicate LO phonons and $L_-$ modes.

**Fig. 5**. (color online) Variation of coherent oscillations by two pump pulses as a function of interpulse separation in n- (a) and p-GaAs (b). The open and filled diamonds indicate LO phonons and $L_-$ modes, respectively. The dashed lines represent fits to a damped harmonic function. The dash-dotted lines show the lifetime for single-pump excitation. The solid lines are calculated using the model of Eq. (1-6).

**Fig. 6**. (color online) Schematic sketch of the kinetic processes involved in coherent control experiments.

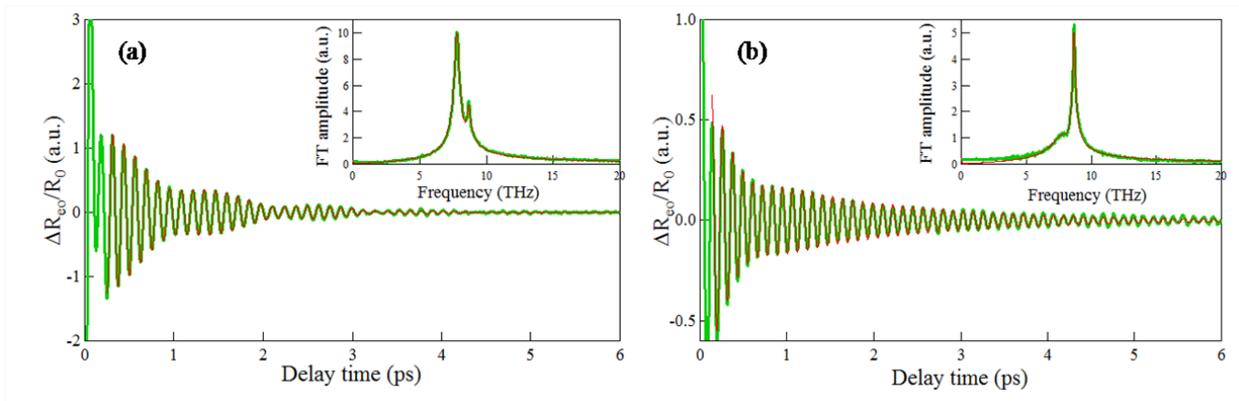

**Fig. 1**.**Hu** *et al*.

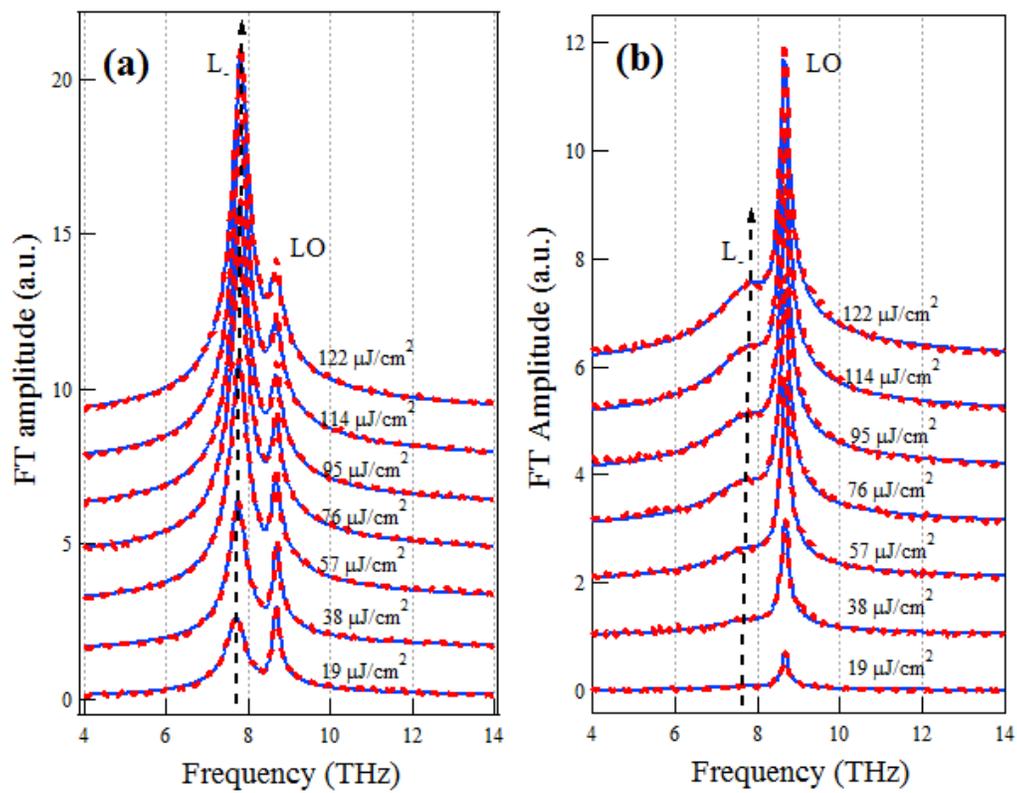

**Fig. 2**. **Hu** *et al*.

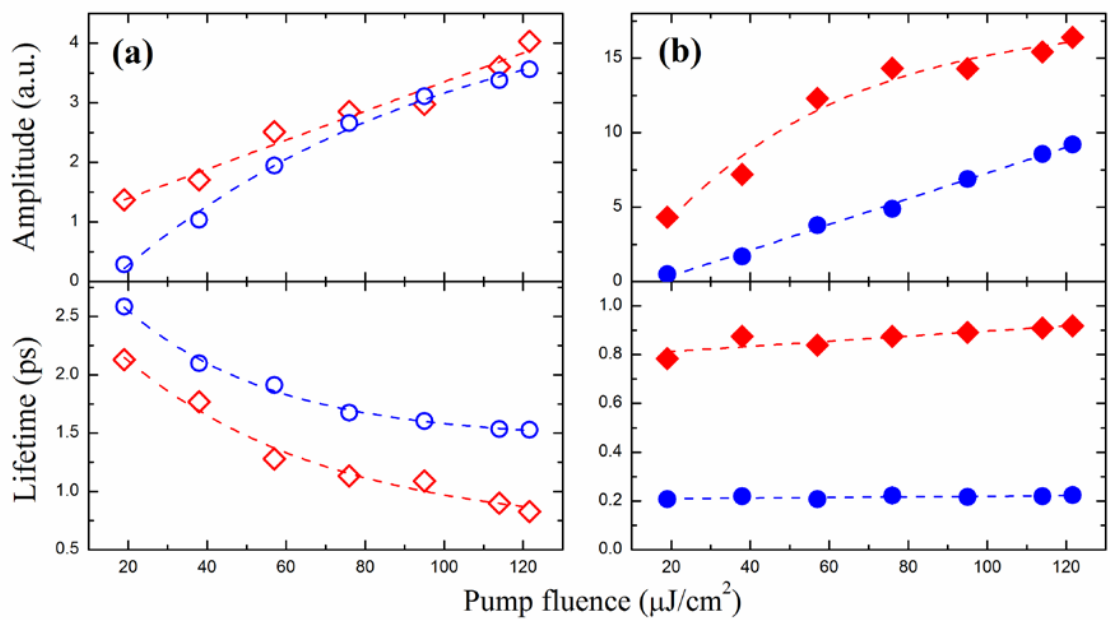

**Fig. 3**. **Hu** *et al*.

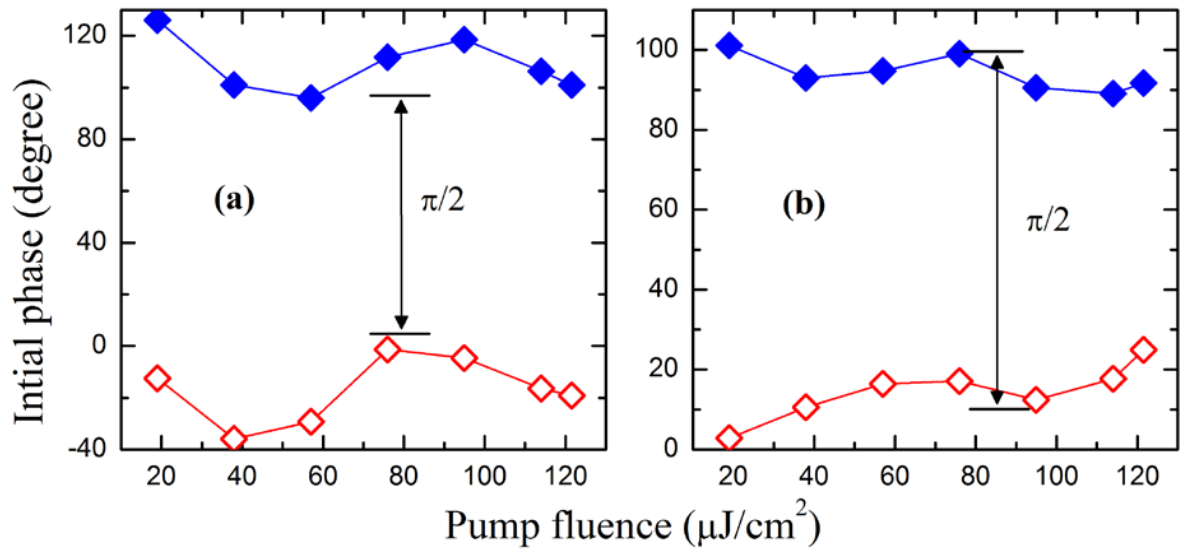

**Fig. 4**. **Hu** *et al*.

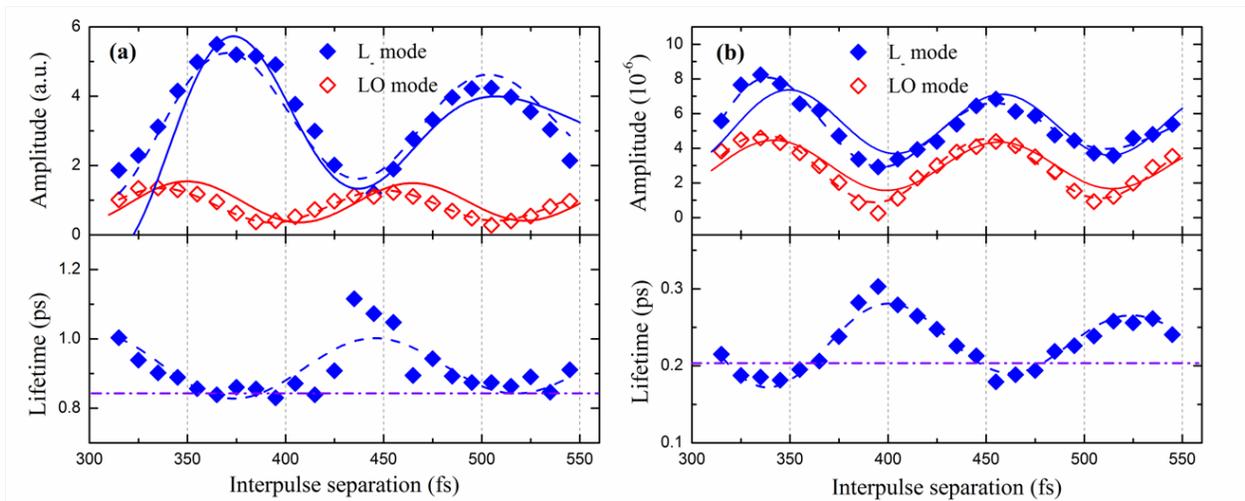

**Fig. 5**. **Hu** *et al*.

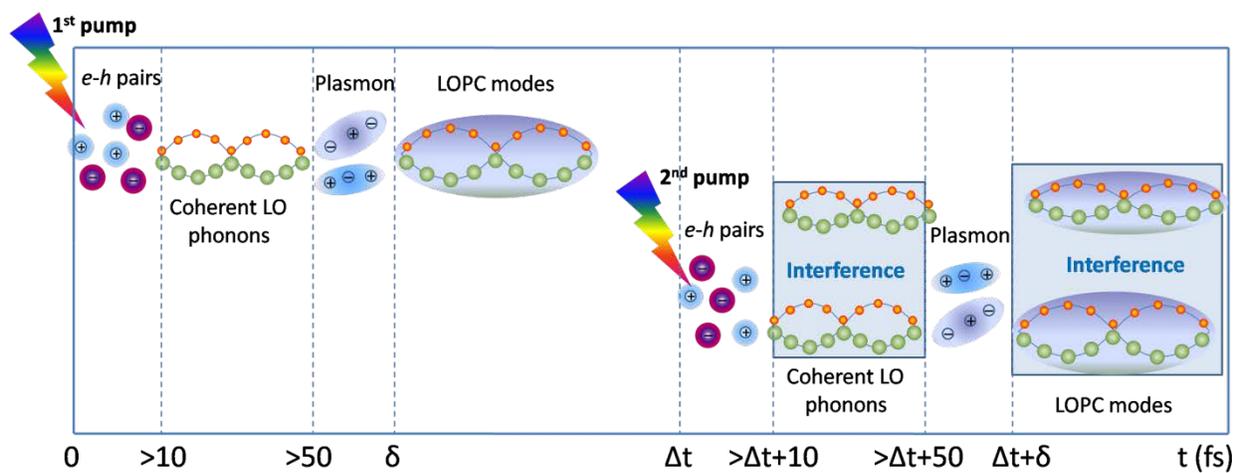

**Fig. 6**. **Hu** *et al*.